\documentclass[onecolumn,floatfix]{revtex4}%
\usepackage{dcolumn}
\usepackage{amsmath}
\usepackage{amssymb}
\usepackage{graphicx}
\usepackage{bm}
\usepackage[T1]{fontenc}
\usepackage{color}
\usepackage{url}
\usepackage[bf]{subfigure}
\usepackage{rotating}
\usepackage{mathrsfs}
\usepackage{multirow}
\usepackage{longtable}
\usepackage{url}
\usepackage{amsthm}
\usepackage{amsfonts}%
\providecommand{\U}[1]{\protect\rule{.1in}{.1in}}
\begin{document}
\date{}
%\title{System performance metrics on the intra- and inter-clusters coupling balance}
\title{Impact of intra and inter-cluster coupling balance on the performance of nonlinear networked systems}

\author{Jiachen Ye$^{1,2,3}$, Peng Ji$^{1,2,3}$}
 \email{pengji@fudan.edu.cn}

\author{David Waxman$^{1,2}$, Wei Lin$^{1,2,3}$, Yamir Moreno$^{4,5,6}$}
 \email{yamir.moreno@gmail.com}

\affiliation{$^{1}$Institute of Science and Technology for Brain-Inspired Intelligence, Fudan University, Shanghai 200433, China}
\affiliation{$^2$LCNBI and LMNS (Fudan University), Ministry of Education, Shanghai 200433, China}
\affiliation{$^{3}$Research Institute of Intelligent and Complex Systems, Fudan University, Shanghai 200433, China}
\affiliation{$^{4}$Institute for Biocomputation and Physics of Complex Systems (BIFI), University of Zaragoza, 50018 Zaragoza, Spain}
\affiliation{$^{5}$Department of Theoretical Physics, University of Zaragoza, 50018 Zaragoza, Spain}
\affiliation{$^{6}$ISI Foundation, Via Chisola 5, 10126 Torino, Italy}

\begin{abstract}
The dynamical and structural aspects of cluster synchronization (CS) in complex systems have been intensively investigated in recent years. Here, we study CS of dynamical systems with intra- and inter-cluster couplings. We propose new metrics that describe the performance of such systems and evaluate them as a function of the strength of the couplings within and between clusters. We obtain analytical results that indicate that spectral differences between the Laplacian matrices associated with the partition between intra- and inter-couplings directly affect the proposed metrics of system performance. Our results show that the dynamics of the system might exhibit an optimal balance that optimizes its performance. Our work provides new insights into the way specific symmetry properties relate to collective behavior, and could lead to new forms to increase the controllability of complex systems and to optimize their stability.

\end{abstract}
\maketitle

\renewcommand{\abstractname }

\section{Introduction}

The relationship between the structure of networks and the dynamics of the systems they represent plays a key
role in a variety of collective phenomena exhibited by natural and engineered systems
\citep{strogatz2001exploring,newman2006structure,boccaletti2006complex,rodrigues2016kuramoto,boccaletti2016explosive}.
Of particular interest is the observation that in many systems patterns that correspond to synchronized
clusters emerge. This phenomenon, known as \textit{cluster synchronization} (CS), is a
widespread (and characteristic) illustration of intra-cluster coherence and
inter-clusters incoherence
\citep{sorrentino2016complete,menara2019stability,cho2017stable,pecora2014cluster}. The understanding of the characteristics of CS is of key relevance, as it has been argued that this phenomenon is of central importance for the proper functioning of
nonlinear systems that have evolved or been designed, such as the human brain
\citep{bullmore2009complex,sporns2013structure,zhou2006hierarchical,kim2018role}
and power grids
\citep{PhysRevLett.109.064101,dorfler2014synchronization,menck2014dead,Yang:2017gh}. Despite several attempts, it is not year clear whether CS will occur and how to identify or predict in advance its emergence.

On the one hand, a considerable amount of prior work has been devoted to the issue of establishing a compact representation of the relationship between the structure and the dynamics\citep{sorrentino2016complete,menara2019stability,zhang2017incoherence,whalen2015observability,nicosia2013remote,golubitsky2012singularities} in systems that display CS. Such a representation facilitates understanding the mechanisms that eventually produce cluster synchronization. For instance, it has been observed that underlying structural symmetries can induce patterns of CS. Interestingly, the reverse is also true, namely, CS can reveal underlying symmetries \citep{nicosia2013remote,pecora2014cluster}. Patterns of CS have also been shown, both experimentally and theoretically, to be induced by modulating structures and by heterogeneous time-delayed couplings \citep{fu2013topological,williams2013experimental}.

On the other hand, and leaving aside the identification of numerous types of emergent CS patterns, the focus has recently been placed in studying the persistence of CS. Group theory, for example, uses the connection between symmetries and nonlinear performance measures to get new insights into the dynamical behavior of both simple \citep{golubitsky2012singularities} and arbitrarily complex networks
\citep{pecora2014cluster}. Indeed, applying group theory to
dynamically equivalent networks facilitates the detection of cluster
synchronization patterns \citep{sorrentino2016complete}. Additionally, both
the degree of cluster symmetry and the spatial distribution of coupling strengths are key factors for the stability of
CS. Admittedly, higher symmetries lead to a reduced region of stability \citep{whalen2015observability}, whereas intra-cluster couplings that
are higher than inter-clusters couplings can induce stronger \textit{local} exponential stability in networks of heterogeneous Kuramoto oscillators
\citep{menara2019stability}. However, to the best of our knowledge, no prior work has investigated the \textit{partitioning} of coupling within and between
clusters, and its relation to nonlinear performance measures on realistic networks.

In this work, we are concerned with the synchronization of clusters in a general setting, as quantified by two performance metrics. We address the
effects of the differences between within and between cluster couplings (henceforth called the \textit{balance} between such couplings) on two
performance metrics. We use irreducible group representations to bridge the connection between structural clusters and the nonlinear performance measures, and provide a general theory that is shown to work for the Kuramoto model and an ecological model. The analytical results are consistent, to a good accuracy, with numerical simulations for several combinations of intra- and inter-cluster couplings.

\section{Methodology}

We consider the following classical dynamical equations
\begin{equation}
\dot{\bm{x}}_{i}(t)=\bm{F}(\bm{x}_{i}(t))-\sum_{j=1}^{N}k_{ij}A_{ij}%
\bm{G}(\bm{x}_{i}(t),\bm{x}_{j}(t)),~~~i=1,2,\dots,N, \label{initial}%
\end{equation}
\noindent where $\bm {x}_{i}$ is an $n$-dimensional column vector
characterizing the state of the $i$'th oscillator; $\bm {F}$ represents the
intrinsic dynamics of each oscillator; and $k_{ij}$ quantifies the strength of the coupling between nodes $i$
and $j$. Moreover, $A_{ij}$ are the elements of a symmetric adjacency matrix $A=\{A_{ij}\}$ which encodes the connectivity pattern of the
underlying network, with $A_{ij}$ equal to $1$ if oscillators $i$ and $j$ are connected and $0$ otherwise. Finally, $\bm {G}$ is the output function of adjacency oscillators, and is also an $n$-dimensional column vector. Eq. (\ref{initial})
governs the general dynamics of numerous network-coupled systems and allows, for instance, to establish a connection between network symmetries
and cluster formation \citep{pecora2014cluster}, and to capture how the rules of spatiotemporal signal-propagation depend on a network's topology \citep{hens2019spatiotemporal}.

As it is know, the structure of a complex system often determines many emergent behaviors and the functioning of the system. For the current phenomenon of interest, CS, the relationship structure-dynamics is no less, that is, the underlying topological features of a network play a key role in the emergence of cluster synchronization. Based on group theory, we can identify symmetries of a network with $N$ nodes and further partition nodes into $M$  clusters, where nodes within the same cluster have identical dynamical behavior \citep{pecora2014cluster}. For notational convenience, we use
$C_{m}$ ($m=1,2,...,M$) to denote the set of nodes in the $m$'th cluster, with all nodes in $C_{m}$ having identical states (i.e., identical $\bm{x}_{i}$)
that are given by $\mathbf{s}_{m}(t)$ and which correspond to synchronous motion. We introduce  $\alpha(i)$, within the range of $[1, M]$, which maps node $i$ onto its corresponding cluster.

We impose small perturbations on the state of each oscillator, which corresponds to a small deviation away from the global state of $M$ synchronized clusters. If $\delta\bm{x}_{i}$ is the perturbation of the state of the $i$'th oscillator, we have $\bm{x}_{i}=\bm{s}_{\alpha(i)}+\delta\bm{x}_{i}$. We
define $\delta\bm {x}=[\delta\bm {x}_{1}^{T},\delta\bm {x}_{2}^{T}%
,\ldots,\delta\bm {x}_{N}^{T}]^{T}$, which is an $n$-dimensional column
vector that contains all perturbations. The corresponding linearized equation
of the perturbations is
\begin{equation}
\delta\dot{\bm {x}}=[D\bm {F}(\bm {s})-D\bm {G}(\bm {s},\bm {s})]\delta
\bm {x}, \label{linearization}%
\end{equation}
where $D\bm{F}(\bm{s})=\text{diag}~[D\bm{F}(\bm{s}_{\alpha(1)}%
),D\bm{F}(\bm{s}_{\alpha(2)}),\dots,D\bm{F}(\bm{s}_{\alpha(N)})~]$, $D\bm{F}$
is the $n\times n$ Jacobian matrix of $\bm{F}$, and $D\bm{G}(\bm{s},\bm{s})$ is
given by
\begin{equation}
D\bm{G}(\bm{s},\bm{s})_{ij}=\left\{
\begin{aligned} &\sum_{j=1}^N k_{ij}A_{ij}D\bm{G}_1(\bm{s}_{\alpha(i)}, \bm{s}_{\alpha(j)}),~~~&i=j,\\ &k_{ij}A_{ij}D\bm{G}_2(\bm{s}_{\alpha(i)}, \bm{s}_{\alpha(j)}),~~~&i\neq j, \end{aligned}\right. \label{Jacobian_original}
\end{equation}

\noindent while $D\bm{G}_{1}$ and $D\bm{G}_{2}$ are, respectively, the first
and last $n$ columns of the $n\times2n$ Jacobian matrix of $\bm{G}$.

Let us now introduce a coherency metric, $H$, which represents the energy
expended when the system relaxes back to its stable state. In terms of a
quadratic cost of phase differences between any pair of connecting nodes
(following \citep{poolla2017optimal}), the metric $H$ is given by
\begin{equation}
H=\int_{0}^{\infty}\sum_{l=1}^{n}\sum_{i,j=1}^{N}A_{ij}[\delta\bm {x}_{i}%
^{l}(t)-\delta\bm {x}_{j}^{l}(t)]^{2}\mathrm{d}t=2\sum_{l=1}^{n}\int%
_{0}^{\infty}\delta\bm {x}^{l^{T}}(t)~L~\delta\bm {x}^{l}(t)\mathrm{d}%
t\text{,} \label{energy_consumption}%
\end{equation}
\noindent where $\delta\bm {x}_{i}^{l}$ is the $l$'th component of
$\delta\bm {x}_{i}$ and $\delta\bm {x}^{l}=[\delta\bm {x}_{1}^{l}%
,\delta\bm {x}_{2}^{l},\ldots,\delta\bm {x}_{N}^{l}]^{T}$.  $L$ is the
Laplacian matrix associated with the adjacency matrix $A$, and it is
defined as $L=D-A$ where $D$ is the diagonal matrix whose elements are the nodes' degree.

The coherency metric, $H$, combines intra- and inter-clusters' interaction and separation, but this combination is hidden in the underlying structure. In order to get a deeper insight into the different contributions to $H$, we introduce another performance metric, henceforth denoted by $J$, which is based on a simple 2 norm that captures the phase variance of the whole system. Therefore, we define
\begin{equation}
J=\int_{0}^{\infty}\Vert\delta\bm {x}(t)\Vert_{2}^{2}\mathrm{d}t=\int%
_{0}^{\infty}\delta\bm {x}^{T}(t)\delta\bm {x}(t)\mathrm{d}t\text{.}
\label{metric J}%
\end{equation}

Let us now define a new coordinate system. To this end, we capitalize on some studies that have found that a unitary matrix $T$, which depends on $A$,  provides a powerful way to transform the linearized equation, Eq.(\ref{linearization}), into a convenient new coordinate system. In this new
coordinate system, the transformed coupling matrix $B=TAT^{-1}$ has a block diagonal form, reflecting the symmetry structure and revealing the hidden
clusters' interaction and separation \citep{pecora2014cluster}. Specifically, the upper-left block of $B$ is an $M\times M$ matrix that describes the
dynamics within the synchronization manifold. The remaining diagonal blocks describe motion transverse to this manifold. Applying $T$ to Eq.
(\ref{linearization}), we rewrite this linearized equation as
\begin{equation}
\dot{\bm {\eta }}=\mathcal{T}%
[D\bm {F}(\bm {s})-D\bm {G}(\bm {s},\bm {s})]\mathcal{T}^{-1}%
\bm {\eta }\text{,}%
\end{equation}
\noindent where $\mathcal{T}=T\bigotimes I_{n}$  and
$\bm {\eta }=\mathcal{T}\delta\bm {x}$. Based on the new coordinate system,
Eq. (\ref{metric J}) can be rewritten as
\begin{equation}
J=\int_{0}^{\infty}\left[  \mathcal{T}\delta\bm {x}(t)\right]  ^{T}\left[
\mathcal{T}\delta\bm {x}(t)\right]  \mathrm{d}t=\int_{0}^{\infty}\bm\eta
^{T}(t)\bm\eta(t)\mathrm{d}t\text{.}%
\end{equation}
\noindent Denoting the first $Mn$ and last $(N-M)n$ exponents of $\bm {\eta }$
by $\bm\eta_{+}$ and $\bm\eta_{-}$, respectively, we divide $J$ into
\begin{equation}
J_{+}=\int_{0}^{\infty}\bm\eta_{+}^{T}(t)\bm\eta_{+}(t)\mathrm{d}%
t\text{\ \ and\ \ }J_{-}=\int_{0}^{\infty}\bm\eta_{-}^{T}(t)\bm\eta
_{-}(t)\mathrm{d}t\text{.}%
\end{equation}

\noindent$J_{+}$ and $J_{-}$ sum the intra-clusters integration and intra-clusters separation, respectively, across clusters. While $J$ reveals more
details of the hidden intra- and inter-cluster combinations, both the coherency metric $H$ and the transformed metric $J$ capture the system stability but from different perspectives. Note that small values of both metrics represent high levels of robustness of the system against disturbances.

Of further interest is to investigate how a redistribution of the intra- and inter-cluster coupling strengths influence the values of $H$ and $J$. For simplicity, we consider the coupling strength matrix $K_{M\times M}$, where the diagonal elements represent the homogeneous intra-coupling strengths between nodes within the same cluster and the off-diagonal elements stand for the heterogeneous inter-coupling strengths between different clusters. The minimization problem (recall that the smaller the value, the higher the robustness) can be formulated as
\begin{equation}
\mathrm{m}\mathrm{i}\mathrm{n}~H\quad\text{subject to }\varphi(K)=c
\end{equation}
where $\varphi(\cdot)$ is a constraint function on elements of $K$ and
$c$ is a constant. To address the minimization problem, we can
further solve the lagrangian
\begin{equation}
\mathcal{L}(K,\lambda)=H-\lambda\lbrack\varphi(K_{11},\ldots K_{MM}%
)-c]\text{,}%
\end{equation}
and obtain the optimal solution satisfying
\begin{equation}
\left\{
\begin{aligned} \frac{\partial \mathcal{L}}{\partial K_{ij}} & = \frac{\partial {H}}{\partial K_{ij}} - \lambda\frac{\partial \varphi}{\partial K_{ij}}=0,~~~~~i,j=1,2,\dots,M, \\ \frac{\partial \mathcal{L}}{\lambda} & = \varphi(K) - c = 0. \end{aligned}\right.
\end{equation}
The above procedure can also be applied to the optimization solution for
minimizing $J$.

\section{Application to two paradigmatic dynamics}

\subsection{Kuramoto model}

To further analyze the stability of the system with respect to the balance between intra- and inter-couplings, we first consider the classical Kuramoto model,
which is governed by the equations
\begin{equation}
\dot{\theta}_{i}=P_{i}-\sum_{j=1}^{N}k_{ij}A_{ij}\mathrm{s}\mathrm{i}%
\mathrm{n}(\theta_{j}-\theta_{i}),\text{\ \ \ }i=1,2,\ldots,N\text{,}%
\end{equation}
\noindent in which $F(\theta_{i})=P_{i}$, and $G(\theta_{i},\theta
_{j})=\mathrm{s}\mathrm{i}\mathrm{n}(\theta_{j}-\theta_{i})$. When the system operates within the regime of stable synchronization, we can build the corresponding Jacobian matrix and, from Eq. (\ref{Jacobian_original}), get that $DG_{1}%
(\theta_{i},\theta_{j})=-\mathrm{c}\mathrm{o}\mathrm{s}(\theta_{j}-\theta
_{i})\approx-1$ and $DG_{2}(\theta_{i},\theta_{j})=\mathrm{c}\mathrm{o}%
\mathrm{s}(\theta_{j}-\theta_{i})\approx1$.

Regarding the coupling balance, one should (in theory) use the Lagrangian of the problem to determine all elements of the coupling-strength
matrix $K$. Although this problem may be solvable, in general, the results obtained are hard to interpret. For simplicity, we only consider diagonal
elements of $K$ equal to $k_{1}$ (the coupling strength within clusters), and the off-diagonal elements of $K$ equal to $k_{2}$ (the
coupling strength between clusters). After an arbitrary set of disturbances,
$\delta\bm\theta(0)=\bm v$, where $\bm v=[v_{1},v_{2},\ldots,v_{N}]^{T}$
$\in\mathbb{R}^{N}$ represents the magnitudes of the disturbances on nodes,
one can obtain the explicit solution of $\delta\bm\theta$ from Eq. (\ref{linearization}) as
\begin{equation}
\delta\bm\theta=e^{-(k_{1}L_{1}+k_{2}L_{2})t}\bm v=e^{-L_{k}t}\bm v\text{,}
\label{delta_theta_explicit}%
\end{equation}
\noindent where $L_{1}$ and $L_{2}$ are the intra-cluster and inter-cluster parts of the Laplacian matrix $L_{k}$, and $L_{k}=k_{1}L_{1}+k_{2}L_{2}$. Here,
$L_{1}=D^{(1)}-A^{(1)}$ and $L_{2}=D^{(2)}-A^{(2)}$. Specifically, $D^{(1)}=\mathrm{d}\mathrm{i}\mathrm{a}\mathrm{g}\{d_{i}^{(1)}\}$, where
$d_{i}^{(1)}$ counts the number of intra-clusters links connecting $i$ within the same cluster $C_{\alpha(i)}$, and $A^{(1)}=\{A_{ij}^{(1)}\}$, with
$A_{ij}^{(1)}$ representing intra-clusters links between nodes $i$ and $j$ within $C_{\alpha(i)}$. $D^{(2)}$ is defined in the same way but for inter-cluster connections, that is, $D^{(2)}=\mathrm{d}\mathrm{i}\mathrm{a}\mathrm{g}\{d_{i}^{(2)}\}$ counts the number of inter-clusters edges linking
node $i$ to nodes that belong to different clusters, and $A^{(2)}=\{A_{ij}^{(2)}\}$ represents inter-cluster edges.

Given the explicit solution of $\delta\bm  \theta$, we can calculate the coherency metric as
\begin{equation}
H =2 \int_{0}^{\infty}\delta\bm \theta^{T} (t) L \delta\bm \theta(t) d t =
\bm v^{T} \left(  \sum_{i =2}^{N}\frac{1}{\lambda_{i}} u_{i} u_{i}^{T}\right)
L \bm v, \label{H_explicit}%
\end{equation}
\noindent(see S2 for the detailed calculation) where $\lambda_{i}$ and $u_{i}$ ($i =2 ,3,\ldots,N$) are the eigenvalues of $L_{k}$ from the smallest to the largest except for $\lambda_{1} =0$ and the corresponding eigenvectors, respectively.

We implement the constraint $k_{1}+k_{2}=1$, that forces a partition of couplings, and allows the investigation of the effects of the intra and inter-coupling balance on the coherency metric $H$. We proceed by obtaining the first and the second derivatives of $H$ with respect to $k_{1}$ (see S2), which leads to
\begin{equation}
\left\{
\begin{array}
[c]{c}%
\frac{dH(k_{1})}{dk_{1}}=-\bm v^{T}\left(  \sum_{i=2}^{N}\frac{1}{\lambda
_{i}^{2}}u_{i}u_{i}^{T}\right)  (L_{1}-L_{2})L\bm v\text{,}\\
\frac{d^{2}H(k_{1})}{dk_{1}^{2}}=2\bm v^{T}\left(  \sum_{i=2}^{N}\frac
{1}{\lambda_{i}^{3}}u_{i}u_{i}^{T}\right)  (L_{1}-L_{2})^{2}L\bm v\text{.}%
\end{array}
\right.  \label{H_derivative}%
\end{equation}

\noindent Equations (\ref{H_derivative}) constitute the theoretical solution for $H$ as a function of $k_{1}$, being the $\lambda_{i}$ the eigenvalues of the matrix $L_{k}=k_{1}(L_{1}-L_{2})+L_{2}$. By varying $k_{1}$ (for instance, increasing it), one can explore how the interplay of the spectra of $L_{1}-L_{2}$ and $L_{2}$ might lead to non-trivial phenomena. Of particular interest, as noted before, is the set of parameters that optimize the system's stability. This can be represented as
\begin{equation}
\mathrm{m}\mathrm{i}\mathrm{n}H\text{ subject to }k_{1}+k_{2}=1\text{ with
}0\leqslant k_{1},k_{2}<1\text{,}%
\end{equation}

Similarly, the explicit solution of state change is%
\begin{equation}
J=\int_{0}^{\infty}\left[  T\delta\bm\theta(t)\right]  ^{T}\left[
T\delta\bm\theta(t)\right]  dt=\frac{1}{2}\bm v^{T}\left(  \sum_{i=2}^{N}%
\frac{1}{\lambda_{i}}u_{i}u_{i}^{T}\right)  \bm v\text{.}%
\end{equation}

\noindent With the matrices%
\begin{equation}
Q_{+}=\left[  I_{M}~|~O_{M\times(N-M)}\right]  ,\text{\ \ \ \ }Q_{-}=\left[
O_{(N-M)\times M}~|~I_{(N-M)}\right]  \text{,}%
\end{equation}
\noindent where $I$ is the identity matrix and $O$ is the zero matrix. The quantities $J_{+}$ and $J_{-}$ can be expressed as%
\begin{align}
J_{+}  &  =\int_{0}^{\infty}\bm\eta_{+}^{T}(t)\bm\eta_{+}(t)dt=\int%
_{0}^{\infty}\delta\bm\theta^{T}(t)T^{T}Q_{+}^{T}Q_{+}T\delta\bm\theta
(t)dt\text{,}\\
J_{-}  &  =\int_{0}^{\infty}\bm\eta_{-}^{T}(t)\bm\eta_{-}(t)dt=\int%
_{0}^{\infty}\delta\bm\theta^{T}(t)T^{T}Q_{-}^{T}Q_{-}T\delta\bm\theta
(t)dt\text{.}%
\end{align}
\newline

\begin{figure}[ptb]
\centering
\includegraphics[width=\linewidth ]{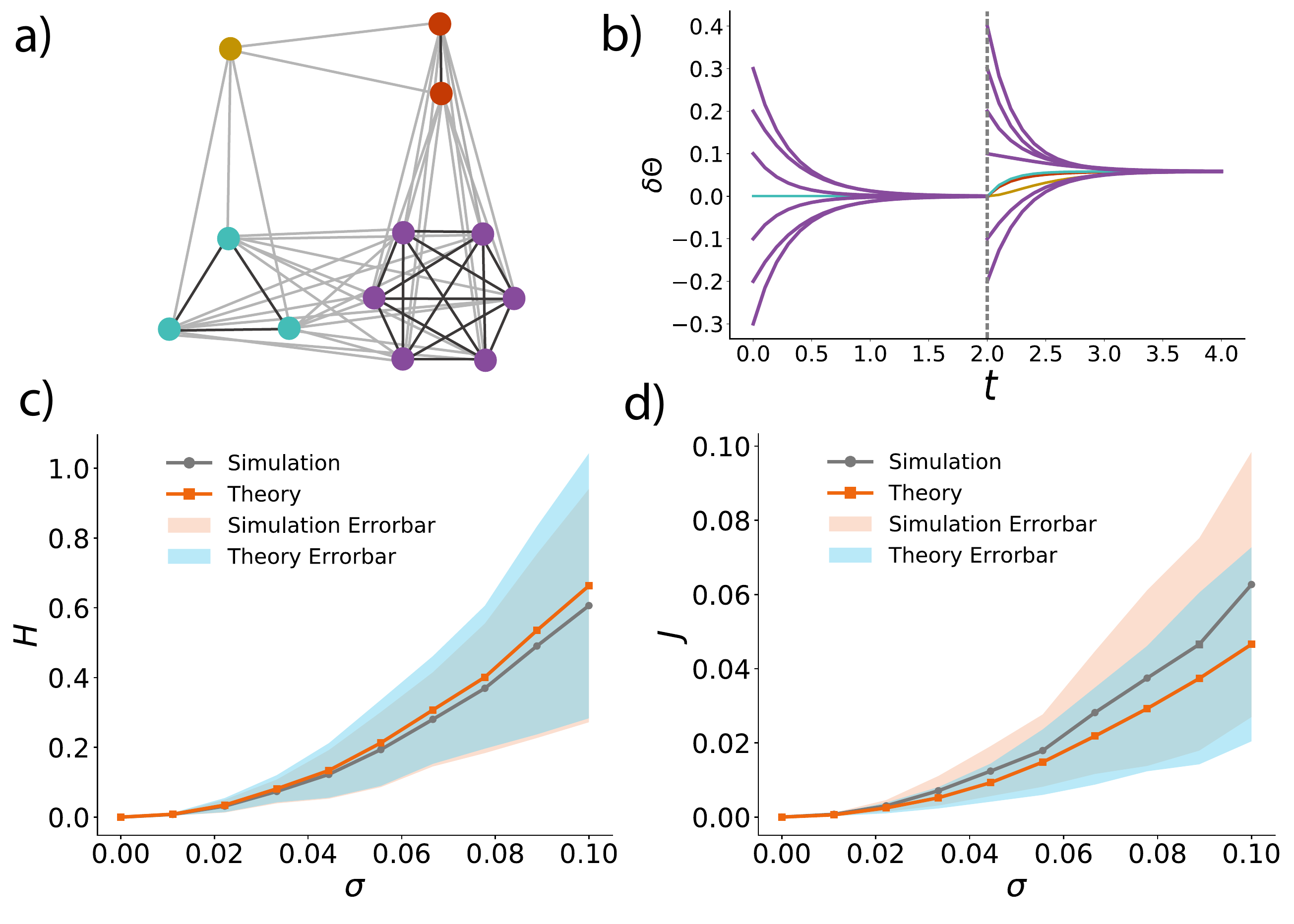}\caption{{\protect\footnotesize Experimental observation of the model network. (a) The model network composed of 12 nodes with color representing cluster partition. (b) Disturbing the purple cluster subject to different conditions ($\sum_{i=7}^{12}v_{i}=0$ at $t=0$ and
$\sum_{i=7}^{12}v_{i}\neq0$ at $t=2$) results in different steady states. (c) Variation between the numerical and the theoretical solutions of $H$ with $\sigma$ for the case of $k_{1}=0.9$ and $k_{2}=0.1$, where $\{v_{i}\}$ obey a normal distribution $N(0,\sigma)$. Each curve corresponds to the average over 100 realizations and the error bars represent the standard deviation. (d) Variation between the numerical and the theoretical solutions of $J$ with $\sigma$ in the same situation.}}%
\end{figure}

In order to check the accuracy of the proposed theoretical framework, we next use a toy network model composed of $12$ nodes. Fig.~1(a) shows the topology of the network, with nodes of the same color belonging to the same cluster. The dark-color edges link nodes within each clusters while the light-color edges link nodes between clusters. If a perturbation of nodes is restricted to be within only one cluster, then we can use one unitary matrix $T$ of the toy model and determine which clusters will be influenced (this depends on the nature of the perturbation). Fig.~1(b) illustrates the time series of each node (with color corresponding to different clusters) after two kinds of perturbations are applied as follows. At $t=0$, we apply perturbations to nodes of the purple cluster with $\sum_{i=7}^{12}v_{i}=0$; purple nodes are affected but other nodes remain unaffected. At $t=2.0$, we again apply perturbations to purple nodes with $\sum_{i=7}^{12}v_{i}\neq0$ and all nodes are affected. After application of the second perturbation, the system approaches a new synchronized state, and the coherency metric and state change quantify the state displacement during this process. Perturbations $\{v_{i}\}$ are in general assumed to obey a normal distribution $N(0,\sigma)$. The strength of perturbations or variations $\sigma$ is crucial to the system stability. Fig.~1(c,d)
illustrates the validation of numerical and theoretical solution of $H$ and $J$ with $\sigma$ given, $k_{1}=0.9$ and $k_{2}=0.1$. With the increase of
$\sigma$, the difference between the numerical and theoretical solution increases progressively as well as the standard deviation.

As noted before, the solutions Eqs. (\ref{H_derivative}) might depend on the interplay/balance between $L_{1}$ and $L_{2}$, which on its turn is determined by how $k_{1}^{\ast}$ is changed. Here, we fix $L_{2}$ and proceed as follows to vary $L_{1}$: i) we increase the connection strength by multiplying by an arbitrary coefficient $\omega$, i.e., $L_{k}=k_{1}\omega L_{1}+k_{2}L_{2}$, and ii) increase the connectivity of $L_{1}$. Fig.~2(a) shows the coherency metric curve with respect to $\omega$. When $\omega$ is relatively small, $H$ increases monotonically with $k_{1}$. However, when $\omega$ is relatively large, $H$ first decreases and then increases with $\omega$. In this case, $H$ has one optimal solution located at $k_{1}^{\ast}$. Moreover, the value of $k_{1}^{\ast}$ increases with $\omega$, an the value of $\omega$ at which there is an optimal solution is larger than the critical point $\omega^{\ast}$, see Fig.~2(b). Note that the critical value $\omega^{\ast}$ satisfies $0=\frac{\partial H(k_{1},\omega)}{\partial k_{1}}|_{k_{1}=\varepsilon,\omega=\omega^{\ast}}$, where $\varepsilon$ is close to 0.

\begin{figure}[ptb]
\centering
\includegraphics[width=\linewidth ]{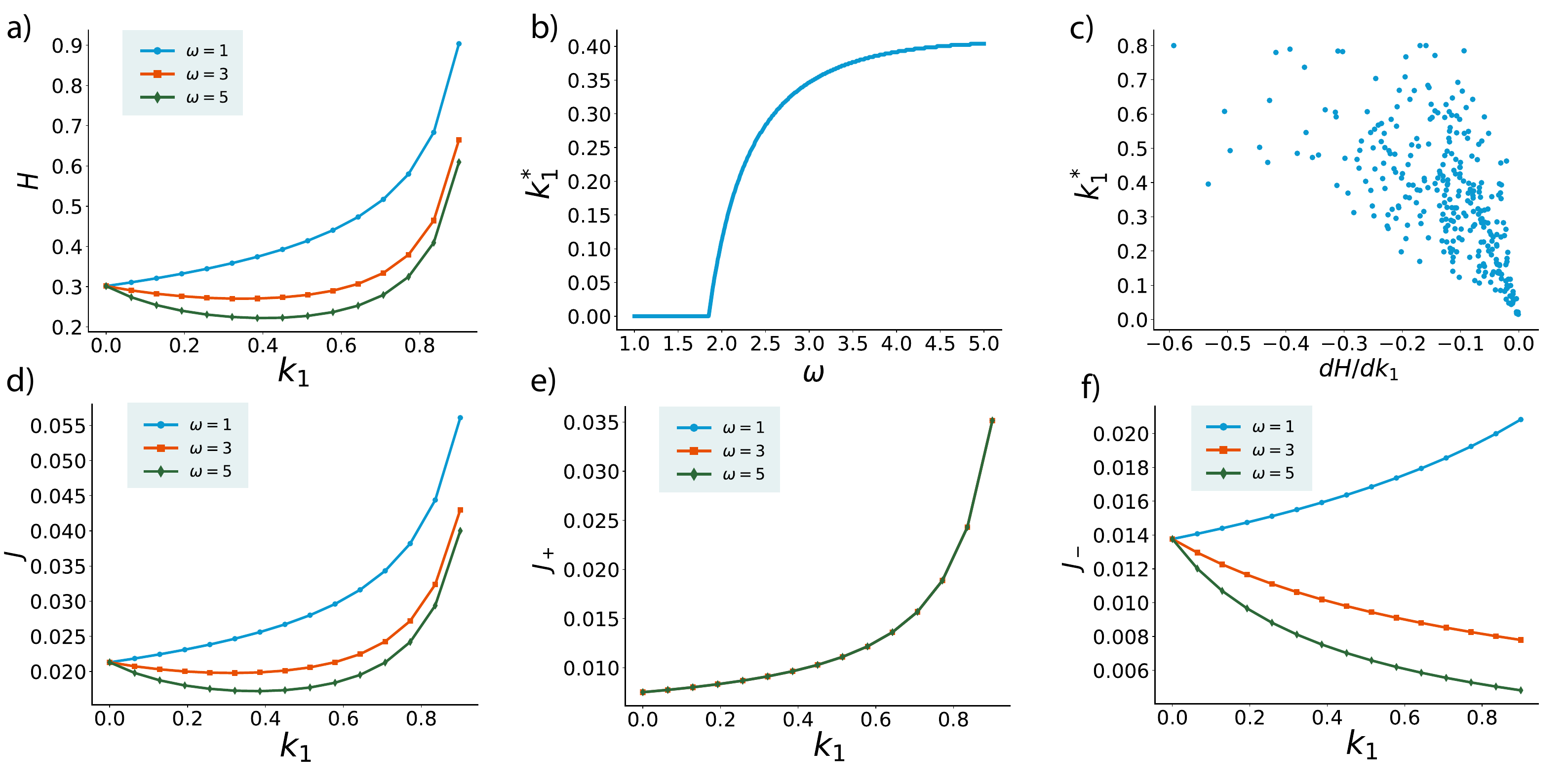}\caption{{\protect\footnotesize Coherency metric and state change with $\omega=1$, $\omega=3$ and $\omega=5$ for a given ${\ \bm v}$. (a) Coherency metric $H$ with different $\omega$. (b) Change of the minimum $k_{1}^{\ast}$ with $\omega$ ranging from 1 to 5. (c) Relation between $k_{1}^{\ast}$ and $\frac{dH(k_{1})}{dk_{1}}\bigg{|}_{k_{1}=\varepsilon}$ (set $\varepsilon=0.01$) with $\omega=5$ and $\{v_{i}\}$ obeying the normal distribution $N(0,0.1)$ for 300 realizations. (d) State change $J$ with different values of $\omega$. (e) Intra-cluster state change $J_{+}$ with different $\omega$s. (f) Inter-cluster state change $J_{-}$ with different  $\omega$.}}%
\end{figure}

Fig.~2(c) indicates that when $\frac{dH(k_{1})}{dk_{1}}|_{k_{1}=\varepsilon}<0$, the smaller the slope of $H(\varepsilon)$, the closer $k_{1}^{\ast
}$ is to 0. We also calculate $J$ when $\omega$ is varied. The explicit solutions of $H$ and $J$, discussed above, indicate that there is only the difference of a constant in the Laplacian matrix between them. Thus, they share the same patterns, as illustrated in Fig.~2(d). The metric $J$ corresponds to global
properties of the whole system and consists of the intra-clusters integration across clusters $J_{+}$ and of the intra-clusters separation across clusters $J_{-}$. Further observation of $J_{+}$ and $J_{-}$ reveals that the inter-cluster part is affected by different $\omega$s, as illustrated in Fig.~2(e) and Fig.~2(f), due to the extra weight added to $L_{1}$ (the inter-cluster part of the Laplacian matrix). This also implies that the dynamics between and within clusters are, in a sense, separated.

\begin{figure}[ptb]
\centering
\includegraphics[width=0.8\linewidth ]{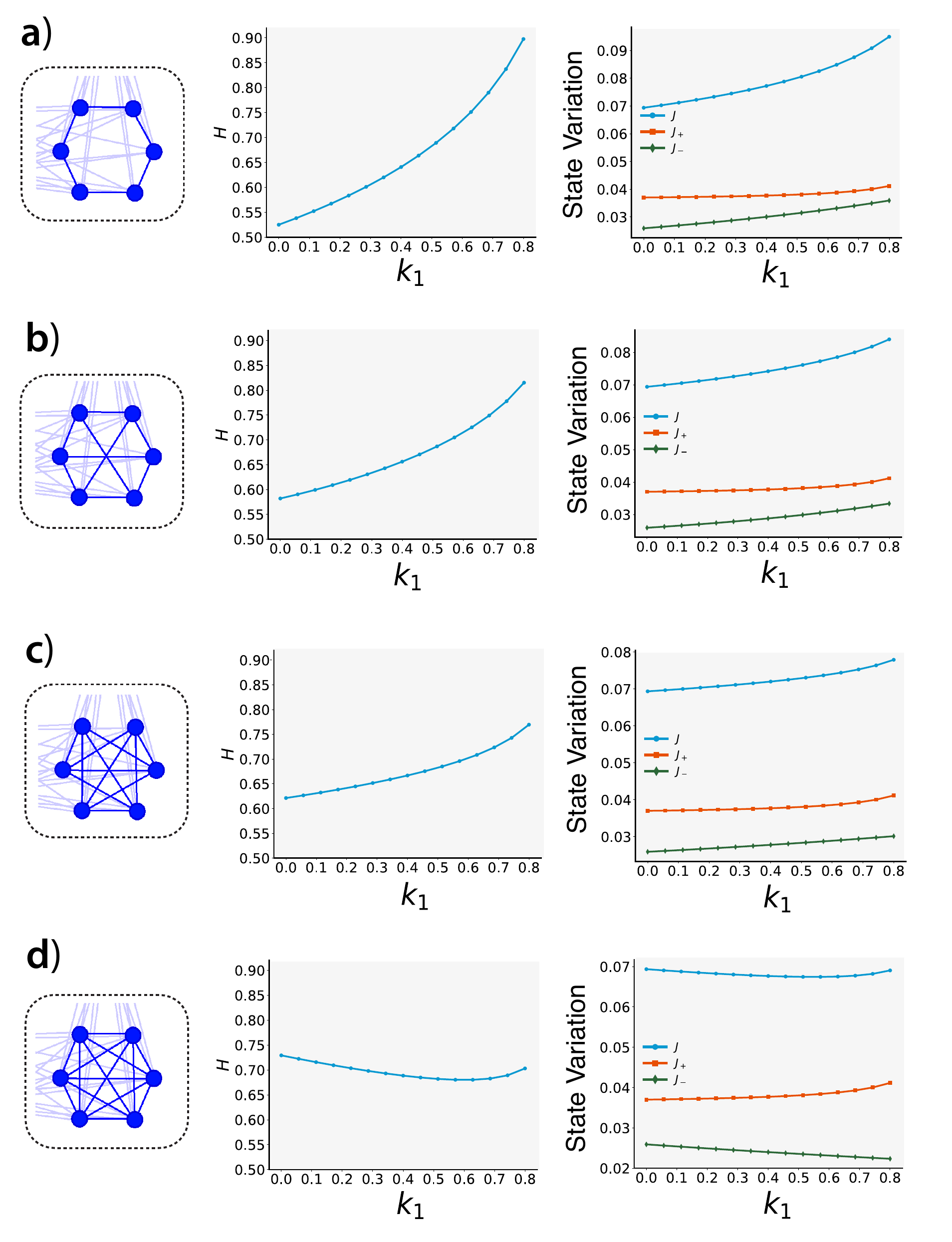}\caption{{\protect\footnotesize Coherency metric and state change with different average inter-cluster degrees for the purple cluster in Figure 1 and for a given ${\ \bm v}$. (a) Average inter-cluster degree equals to 2. (b) Average inter-cluster degree equals to 3. (c) Average inter-cluster degree equals to 4. (d) Average inter-cluster degree equals to 5.}}%
\end{figure}

In addition to the connection strength, we have also varied the connectivity of the network. Results are shown in Fig.~3. In particular, we find different coherency metric curves as well as state change curves with respect to different average degrees of the perturbed (purple) cluster. When the average degree is relatively small, both $H$ and $J$ increase monotonically with $k_{1}$. However, when this average degree is relatively large (the fully connected
network in Fig.~3(d)), $H$ and $J$ exhibit non-trivial solutions with a minimum at $k_{1}^{\ast}$. This situation is similar to that observed in Fig. 2 for high values of $\omega$.

\subsection{Dynamics of Mutualism networks}

In addition to the Kuramoto model, we also consider another paradigmatic dynamics corresponding to a real system, e.g., that of mutualistic interactions among species in an ecological network. We consider the following equations \citep{Gao2016Universal} to describe the evolution of the number of individuals, or abundance, of species $i$, $x_{i}(t)$,
\begin{equation}
\dot{x}_{i}=B_{i}+r_{i}x_{i}\left(  1-\frac{x_{i}}{C_{i}}\right)  \left(
\frac{x_{i}}{G_{i}}-1\right)  +\sum_{j=1}^{N}A_{ij}\frac{x_{i}x_{j}}%
{D_{i}+E_{i}x_{i}+H_{j}x_{j}},\label{mutualism}%
\end{equation}
where, on the right hand side of the equation, the first term, $B_{i}$, captures the incoming migration rate of $i$ from neighboring ecosystems; the
second term accounts for the system's logistic growth with a carrying capacity $C_{i}$, and the Allee effect indicates that, for low population ($x_{i}<G_{i}<C_{i}$), the population size of species $i$ decreases; the third term encodes the mutualistic dynamics, which is modulated by the mutualistic interactions $(i,j)$ given by the matrix $A$. Here, we use symbiotic interactions $A_{ij}$ constructed from plant-pollinator relationships as a classic kind of mutualistic relationships. Plants need pollinators to reproduce and pollinators feed mainly on nectar from plants.

In this system, the abundance $x_{i}$ corresponds to one species $i$ or cluster. Therefore, the second term quantifies the intra-species influence and
the third term accounts for inter-species relations. Based on this system, we aim to quantify the balance of intra- and inter-cluster effects on the stability of the system. With this goal in mind, we additionally include the intra-species coupling strength $k_{1}$ in the second term of the above system of equations and the inter-species coupling strength $k_{2}$ to its third term. The additional coupling strengths $k_{1}$ and $k_{2}$ could account for exogenous factors with the capability of impacting the abundance of species in the system. For instance, favorable environmental conditions might create a better scenario in which pollinators and plants reproduce more. This would correspond to an increase of the intra-species coupling strength $k_{1}$. We also note that the same conditions that favor the increase of $k_{1}$ might imply a reduction of $k_{2}$. Thus, the final equations are
\begin{equation}
\dot{x}_{i}=B_{i}+k_{1}r_{i}x_{i}\left(  1-\frac{x_{i}}{C_{i}}\right)  \left(
\frac{x_{i}}{G_{i}}-1\right)  +k_{2}\sum_{j=1}^{N}A_{ij}\frac{x_{i}x_{j}%
}{D_{i}+E_{i}x_{i}+H_{j}x_{j}}.\label{mutualism_2}%
\end{equation}

We shall investigate the system stability by adjusting the balance between $k_{1}$ and $k_{2}$. Here, following the above procedure, we have the constraint $k_{1}+k_{2}=1$. Moreover, the underlying species interactions accounts for ecological interactions that are obtained from the web of life project. Specifically, each dataset is represented by a rectangular matrix $M$, with $M_{ij}$ representing the mutualistic relationship between plant $i$ and pollinator $j$. We construct the adjacency matrix $A$ as%
\begin{equation}
A=\left[
\begin{matrix}
0   & M\\
M^T & 0 
\end{matrix}
\right].
\end{equation}
\noindent In other words, $A$ represents interactions between different species (plants
and pollinators) and competitive interactions between plants and pollinators are not given by the interaction matrix. However, if one projects links
between the plants as the edges connected by pollinators, it is possible to define the pollinators' projection network. The $(i,j)$ element of the
corresponding adjacency matrix $C_{po}$ equals 1 if pollinator $i$ and pollinator $j$ pollinate the same plant, or equals to 0 otherwise. Similarly, one can also define plants' projection links of the corresponding adjacency matrix $C_{pl}$. Altogether, the system of interactions can be considered as a
two-layer network, whose sketch map is shown in Fig.4(a). The corresponding adjacency matrix is%
\begin{equation}
\tilde{A}=\left[
\begin{matrix}
C_{p\ell}   & M\\
M^T & C_{po}
\end{matrix}
\right].
\end{equation}
To investigate the balance with respect to intra- and inter-species coupling, we follow the above process. Specifically, we numerically integrate Eq.
(\ref{mutualism_2}) and consider the following nonlinear programming problem%
\begin{equation}
\mathrm{m}\mathrm{i}\mathrm{n}\text{ }H\text{ subject to }k_{1}+k_{2}=1\text{
with }0\leqslant k_{1},k_{2}<1.
\end{equation}

\noindent The same process can also be followed for $J$. Note that here each node represents one species (cluster), and therein $J=J_{+}$ and $J_{-}=0$.

\begin{figure}[ptb]
\centering
\includegraphics[width=0.9\linewidth ]{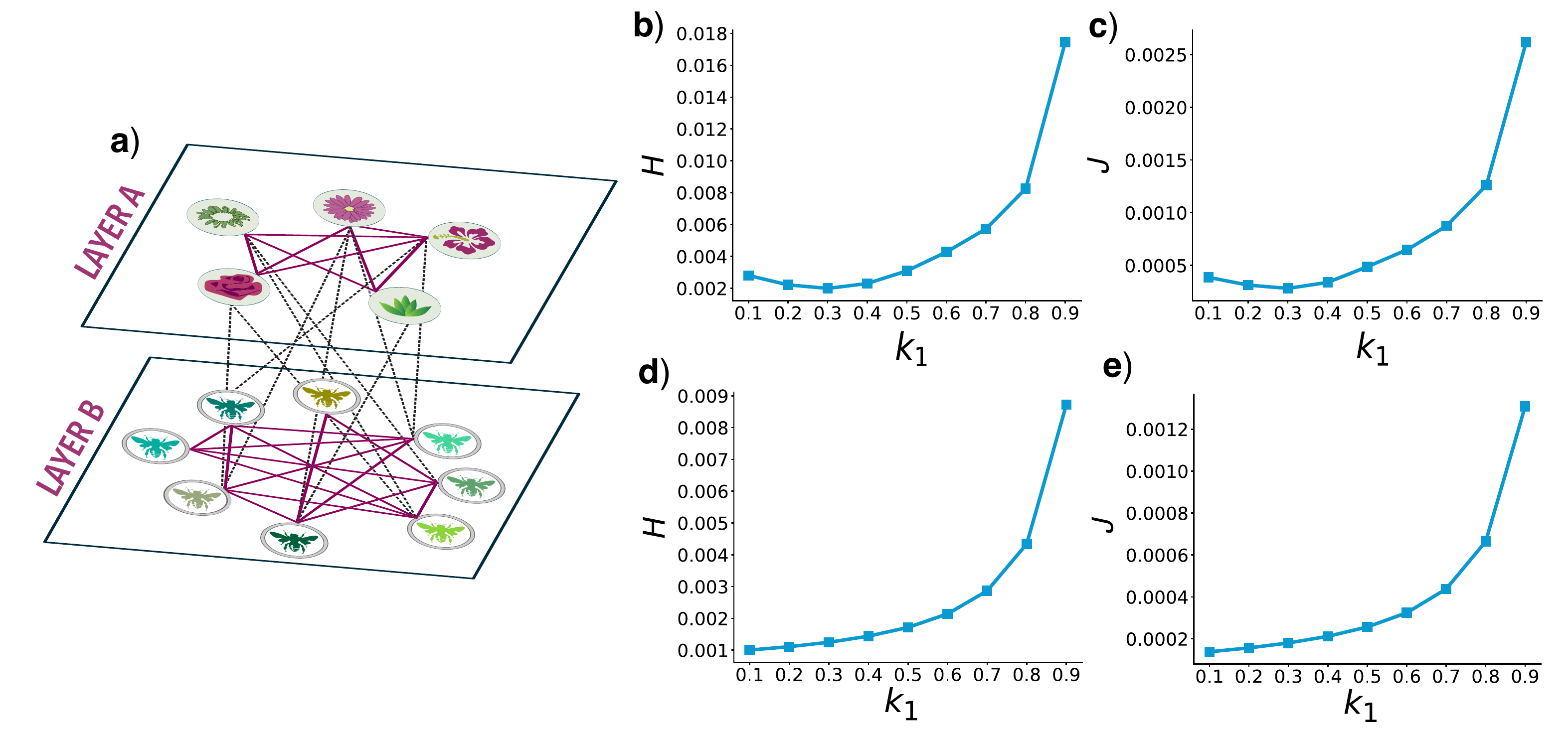}\caption{{\protect\footnotesize The figure shows a schematic representation of a mutualistic network (panel a) and results for the  coherency and state change metrics for two different mutualism networks (panels (b through e). We have set $B_{i}=B=0.01$, $r_{i}=r=0.01$, $C_{i}=C=5$, $G_{i}=G=1$, $D_{i}=D=5$, $E_{i}=E=0.9$ and $H_{i}=H=0.1$. ${\bm v}$ obeys a normal distribution $N(0,0.01)$. (b) Coherency metric for a network composed of 16 plants and 44 pollinators with 278 mutualistic interactions (network MPL46). (c) State change for the same network used in panel b. (d) Coherency metric for the network composed of 11 plants and 38 pollinators with 106 mutualistic interactions (network MPL08). (e) State change for the same network used in panel d.}}%
\end{figure}

There are 149 mutualistic networks provided by the web of life project. We have arbitrarily selected some of such networks for our numerical analysis. Results show that, depending on the selected networks, the coherency metric curves could either decrease first and then increase, or increase monotonically as shown in Fig. 5. The phenomena remain consistent for state change $J$. Fig. 5 shows these two limiting results obtained for two networks of the dataset (see S4 for more curves corresponding to different networks).

\section{Conclusions}

Summarizing, in this manuscript we have investigated what is the impact that changes of the balance between intra- and inter-cluster coupling strengths induce on the stability of the system. In particular, we partition nodes into clusters using irreducible representation theory and linearize the system in the region of cluster synchronization. Depending on the nature of perturbations, only one or multiple clusters will be affected. We have proposed and evaluated two different metrics, namely $H$, which describes the energy that the system consumes to get back to a steady state, and $J$, which captures the state variations. The two metrics quantify stability, but from different points of view with respect to the coupling strength. Our results show that for both metrics the system could exhibit nontrivial behavior with variations of the intra- and inter-coupling strengths. The proposed theoretical approach has been applied to analyze two explicit dynamical models, e.g., the Kuramoto model and the dynamics of a mutualistic ecological network. For the first case, we have used a synthetic network, whereas the latter implements realistic systems. Our results could provide new hints in the quest to control the dynamics of networked systems.

\section*{Acknowledgement}

This work has used the Web of Life dataset (www.web-of-life.es). PJ acknowledges National Key R\&D Program of China (2018YFB0904500), Natural Science Foundation of Shanghai, the Program for Professor of Special Appointment (Eastern Scholar) at Shanghai Institutions of Higher Learning and by NSFC 269 (11701096), National Science Foundation of China under Grant 61773125. YM acknowledges partial support from Intesa Sanpaolo Innovation Center, the Government of Aragon, Spain through grant E36-17R (FENOL), and by MINECO and FEDER funds (FIS2017-87519-P).

\section*{Appendix: Details of mathematical derivations and further results.}

\subsubsection*{S1. $\bm{A}$, $\bm{T}$ and $\bm{B}$ of the example}

Using a discrete algebra software, it is straightforward to determine the symmetries of $A$ and the transformation matrix $T$. We show the results
applied to the network showed in Fig. 1(a):%

\begin{equation}
A=\left[
\begin{matrix}
0 & 1 & 1 & 1 & 1 & 1 & 0 & 0 & 0 & 0 & 0 & 0\\
1 & 0 & 1 & 0 & 0 & 0 & 1 & 1 & 1 & 1 & 1 & 1\\
1 & 1 & 0 & 0 & 0 & 0 & 1 & 1 & 1 & 1 & 1 & 1\\
1 & 0 & 0 & 0 & 1 & 1 & 1 & 1 & 1 & 1 & 1 & 1\\
1 & 0 & 0 & 1 & 0 & 1 & 1 & 1 & 1 & 1 & 1 & 1\\
1 & 0 & 0 & 1 & 1 & 0 & 1 & 1 & 1 & 1 & 1 & 1\\
0 & 1 & 1 & 1 & 1 & 1 & 0 & 1 & 0 & 0 & 0 & 1\\
0 & 1 & 1 & 1 & 1 & 1 & 1 & 0 & 1 & 0 & 0 & 0\\
0 & 1 & 1 & 1 & 1 & 1 & 0 & 1 & 0 & 1 & 0 & 0\\
0 & 1 & 1 & 1 & 1 & 1 & 0 & 0 & 1 & 0 & 1 & 0\\
0 & 1 & 1 & 1 & 1 & 1 & 0 & 0 & 0 & 1 & 0 & 1\\
0 & 1 & 1 & 1 & 1 & 1 & 1 & 0 & 0 & 0 & 1 & 0\\
\end{matrix}
\right]  .
\end{equation}

\noindent There is one trivial cluster (\{1\}) and three non-trivial clusters (\{2,3\}, \{4,5,6\} and \{7,8,9,10,11,12\}). The transformation matrix is%

\begin{equation}
T=\left[
\begin{matrix}
0 & 0 & 0 & 0 & 0 & 0 & -\frac{\sqrt{6}}{6} & -\frac{\sqrt{6}}{6} &
-\frac{\sqrt{6}}{6} & -\frac{\sqrt{6}}{6} & -\frac{\sqrt{6}}{6} & -\frac
{\sqrt{6}}{6}\\
1 & 0 & 0 & 0 & 0 & 0 & 0 & 0 & 0 & 0 & 0 & 0\\
0 & -\frac{\sqrt{2}}{2} & -\frac{\sqrt{2}}{2} & 0 & 0 & 0 & 0 & 0 & 0 & 0 &
0 & 0\\
0 & 0 & 0 & -\frac{\sqrt{3}}{3} & -\frac{\sqrt{3}}{3} & -\frac{\sqrt{3}}{3} &
0 & 0 & 0 & 0 & 0 & 0\\
0 & 0 & 0 & 0 & 0 & 0 & -\frac{\sqrt{6}}{6} & \frac{\sqrt{6}}{6} &
-\frac{\sqrt{6}}{6} & \frac{\sqrt{6}}{6} & -\frac{\sqrt{6}}{6} & \frac
{\sqrt{6}}{6}\\
0 & -\frac{\sqrt{2}}{2} & \frac{\sqrt{2}}{2} & 0 & 0 & 0 & 0 & 0 & 0 & 0 & 0 &
0\\
0 & 0 & 0 & -\frac{\sqrt{6}}{3} & \frac{\sqrt{6}}{6} & \frac{\sqrt{6}}{6} &
0 & 0 & 0 & 0 & 0 & 0\\
0 & 0 & 0 & 0 & \frac{\sqrt{2}}{2} & -\frac{\sqrt{2}}{2} & 0 & 0 & 0 & 0 & 0 &
0\\
0 & 0 & 0 & 0 & 0 & 0 & 0 & -\frac{1}{2} & \frac{1}{2} & 0 & -\frac{1}{2} &
\frac{1}{2}\\
0 & 0 & 0 & 0 & 0 & 0 & -\frac{\sqrt{3}}{3} & \frac{\sqrt{3}}{6} & \frac
{\sqrt{3}}{6} & -\frac{\sqrt{3}}{3} & \frac{\sqrt{3}}{6} & \frac{\sqrt{3}}%
{6}\\
0 & 0 & 0 & 0 & 0 & 0 & 0 & -\frac{1}{2} & -\frac{1}{2} & 0 & \frac{1}{2} &
\frac{1}{2}\\
0 & 0 & 0 & 0 & 0 & 0 & -\frac{\sqrt{3}}{3} & -\frac{\sqrt{3}}{6} &
\frac{\sqrt{3}}{6} & \frac{\sqrt{3}}{3} & \frac{\sqrt{3}}{6} & -\frac{\sqrt
{3}}{6}%
\end{matrix}
\right]  ,
\end{equation}

\noindent and the block diagonal coupling matrix is%

\begin{equation}
B=\left[
\begin{matrix}
2 & 0 & 2\sqrt{3} & 3\sqrt{2} & 0 & 0 & 0 & 0 & 0 & 0 & 0 & 0\\
0 & 0 & -\sqrt{2} & -\sqrt{3} & 0 & 0 & 0 & 0 & 0 & 0 & 0 & 0\\
2\sqrt{3} & -\sqrt{2} & 1 & 0 & 0 & 0 & 0 & 0 & 0 & 0 & 0 & 0\\
3\sqrt{2} & -\sqrt{3} & 0 & 2 & 0 & 0 & 0 & 0 & 0 & 0 & 0 & 0\\
0 & 0 & 0 & 0 & -2 & 0 & 0 & 0 & 0 & 0 & 0 & 0\\
0 & 0 & 0 & 0 & 0 & -1 & 0 & 0 & 0 & 0 & 0 & 0\\
0 & 0 & 0 & 0 & 0 & 0 & -1 & 0 & 0 & 0 & 0 & 0\\
0 & 0 & 0 & 0 & 0 & 0 & 0 & -1 & 0 & 0 & 0 & 0\\
0 & 0 & 0 & 0 & 0 & 0 & 0 & 0 & -1 & 0 & 0 & 0\\
0 & 0 & 0 & 0 & 0 & 0 & 0 & 0 & 0 & -1 & 0 & 0\\
0 & 0 & 0 & 0 & 0 & 0 & 0 & 0 & 0 & 0 & 1 & 0\\
0 & 0 & 0 & 0 & 0 & 0 & 0 & 0 & 0 & 0 & 0 & 1\\
\end{matrix}
\right]  .
\end{equation}

\subsubsection*{S2. Detailed proof of the explicit solution of ${\bm{H}}$ and
its derivative}

The key point of the equality in Equation (\ref{H_explicit}) and Equation
(\ref{H_derivative}) lies in whether $L_{1}$ commutes with $L_{2}$. The
Laplacian matrix of the network is
\begin{equation}
L=\left[
\begin{matrix}
5 & -1 & -1 & -1 & -1 & -1 & 0 & 0 & 0 & 0 & 0 & 0\\
-1 & 8 & -1 & 0 & 0 & 0 & -1 & -1 & -1 & -1 & -1 & -1\\
-1 & -1 & 8 & 0 & 0 & 0 & -1 & -1 & -1 & -1 & -1 & -1\\
-1 & 0 & 0 & 9 & -1 & -1 & -1 & -1 & -1 & -1 & -1 & -1\\
-1 & 0 & 0 & -1 & 9 & -1 & -1 & -1 & -1 & -1 & -1 & -1\\
-1 & 0 & 0 & -1 & -1 & 9 & -1 & -1 & -1 & -1 & -1 & -1\\
0 & -1 & -1 & -1 & -1 & -1 & 7 & -1 & 0 & 0 & 0 & -1\\
0 & -1 & -1 & -1 & -1 & -1 & -1 & 7 & -1 & 0 & 0 & 0\\
0 & -1 & -1 & -1 & -1 & -1 & 0 & -1 & 7 & -1 & 0 & 0\\
0 & -1 & -1 & -1 & -1 & -1 & 0 & 0 & -1 & 7 & -1 & 0\\
0 & -1 & -1 & -1 & -1 & -1 & 0 & 0 & 0 & -1 & 7 & -1\\
0 & -1 & -1 & -1 & -1 & -1 & -1 & 0 & 0 & 0 & -1 & 7
\end{matrix}
\right]  .
\end{equation}

\noindent The intra-cluster part is
\begin{equation}
L_{1}=\left[
\begin{matrix}
0 & 0 & 0 & 0 & 0 & 0 & 0 & 0 & 0 & 0 & 0 & 0\\
0 & 1 & -1 & 0 & 0 & 0 & 0 & 0 & 0 & 0 & 0 & 0\\
0 & -1 & 1 & 0 & 0 & 0 & 0 & 0 & 0 & 0 & 0 & 0\\
0 & 0 & 0 & 2 & -1 & -1 & 0 & 0 & 0 & 0 & 0 & 0\\
0 & 0 & 0 & -1 & 2 & -1 & 0 & 0 & 0 & 0 & 0 & 0\\
0 & 0 & 0 & -1 & -1 & 2 & 0 & 0 & 0 & 0 & 0 & 0\\
0 & 0 & 0 & 0 & 0 & 0 & 2 & -1 & 0 & 0 & 0 & -1\\
0 & 0 & 0 & 0 & 0 & 0 & -1 & 2 & -1 & 0 & 0 & 0\\
0 & 0 & 0 & 0 & 0 & 0 & 0 & -1 & 2 & -1 & 0 & 0\\
0 & 0 & 0 & 0 & 0 & 0 & 0 & 0 & -1 & 2 & -1 & 0\\
0 & 0 & 0 & 0 & 0 & 0 & 0 & 0 & 0 & -1 & 2 & -1\\
0 & 0 & 0 & 0 & 0 & 0 & -1 & 0 & 0 & 0 & -1 & 2
\end{matrix}
\right]  ,
\end{equation}

\noindent and the inter-cluster part is
\begin{equation}
L_{2}=\left[
\begin{matrix}
5 & -1 & -1 & -1 & -1 & -1 & 0 & 0 & 0 & 0 & 0 & 0\\
-1 & 7 & 0 & 0 & 0 & 0 & -1 & -1 & -1 & -1 & -1 & -1\\
-1 & 0 & 7 & 0 & 0 & 0 & -1 & -1 & -1 & -1 & -1 & -1\\
-1 & 0 & 0 & 7 & 0 & 0 & -1 & -1 & -1 & -1 & -1 & -1\\
-1 & 0 & 0 & 0 & 7 & 0 & -1 & -1 & -1 & -1 & -1 & -1\\
-1 & 0 & 0 & 0 & 0 & 7 & -1 & -1 & -1 & -1 & -1 & -1\\
0 & -1 & -1 & -1 & -1 & -1 & 5 & 0 & 0 & 0 & 0 & 0\\
0 & -1 & -1 & -1 & -1 & -1 & 0 & 5 & 0 & 0 & 0 & 0\\
0 & -1 & -1 & -1 & -1 & -1 & 0 & 0 & 5 & 0 & 0 & 0\\
0 & -1 & -1 & -1 & -1 & -1 & 0 & 0 & 0 & 5 & 0 & 0\\
0 & -1 & -1 & -1 & -1 & -1 & 0 & 0 & 0 & 0 & 5 & 0\\
0 & -1 & -1 & -1 & -1 & -1 & 0 & 0 & 0 & 0 & 0 & 5
\end{matrix}
\right]  .
\end{equation}

Since two symmetric matrices are commutative if and only if their matrix product is symmetric, we just need to prove that $L_{1}L_{2}$ is symmetric. In
other words, if we let $L_{1} = [\ell_{1}^{(1)},\ell_{2}^{(1)},\dots,\ell_{N}^{(1)}]$ and $L_{2} = [\ell_{1}^{(2)},\ell_{2}^{(2)},\dots,\ell_{N}^{(2)}]$, we need to prove $\ell_{i}^{(1)^{T}}\ell_{j}^{(2)} = \ell_{j}^{(1)^{T}}\ell_{i}^{(2)}$ for any $1 \leqslant i,j \leqslant N$. Consider the following two situations:

\begin{itemize}
\item [i)] Node $i$ and node $j$ belong to different clusters. We set cluster $C_{\alpha(i)} = [i_{1},i_{2},\dots, i_{N_{\alpha(i)}}]$. Because of the
definition of $L_{1}$, only the $i_{1}$th, $i_{2}$th, $\dots$, $i_{N_{\alpha(i)}}$th components of $\ell_{i}^{(1)}$ have nonzero values and their sum
equals 0. The corresponding components of $\ell_{j}^{(2)}$ all equal to $-1$ if cluster $C_{\alpha(i)}$ and cluster $C_{\alpha(j)}$ are connected,
or $0$ otherwise. No matter whether cluster $C_{\alpha(i)}$ and cluster $C_{\alpha(j)}$ are connected, $\ell_{i}^{(1)^{T}}\ell_{j}^{(2)} =0$.
Similarly, we get $\ell_{j}^{(1)^{T}}\ell_{i}^{(2)} =0$ and $\ell_{i}^{(1)^{T}}\ell_{j}^{(2)} = \ell_{j}^{(1)^{T}}\ell_{i}^{(2)}$.

\item [ii)] Node $i$ and node $j$ belong to the same cluster. As mentioned in i), only the $i_{1}$th, $i_{2}$th, $\dots$, $i_{N_{\alpha(i)}}$th components of
$\ell_{i}^{(1)}$ have nonzero values. The corresponding components of $\ell_{j}^{(2)}$ are equal to $d_{j}^{(2)}$ if they are diagonal elements of
$L$, or 0 otherwise. $\ell_{i}^{(1)^{T}}\ell_{j}^{(2)}$ equals to $-d_{j}^{(2)}$ if node $i$ and node $j$ are connected, or 0 otherwise. Similarly, we get the same case of $\ell_{j}^{(1)^{T}}\ell_{i}^{(2)}$ and $\ell_{i}^{(1)^{T}}\ell_{j}^{(2)} = \ell_{j}^{(1)^{T}}\ell_{i}^{(2)}$.

\end{itemize}

To sum up, $L_{1}$ and $L_{2}$ are commutative. $L$ and $L_{k}$ are the linear combination of $L_{1}$ and $L_{2}$ so that every pair of these four matrices is commutative. Next, we prove Equation (\ref{H_explicit}) and Equation (\ref{H_derivative}). Substituting Equation (\ref{delta_theta_explicit}) into
Equation (\ref{energy_consumption}), we obtain%

\begin{equation}
H = 2 \int_{0}^{\infty}\delta\theta^{T} L \delta\theta\mathrm{d}t = 2 \int%
_{0}^{\infty}{\bm v}^{T} e^{-L_{k}t} L e^{-L_{k}t} {\bm v} \mathrm{d}t.
\end{equation}

\noindent The expansion of $e^{-L_{k}t} $ in matrix power series reads %

\begin{equation}
e^{-L_{k}t} = \sum_{n=0} \frac{1}{n!}(-L_{k}t)^{n}=\sum_{n=0} \frac{(-t)^{n}%
}{n!}(k_{1}L_{1}+k_{2}L_{2})^{n}=\sum_{n=0} \frac{(-t)^{n}}{n!} \sum_{m=0}^{n}
(k_{1}L_{1})^{m}(k_{2}L_{2})^{n-m},
\end{equation}
and $e^{-L_{k}t}$ commutes with $L$. This gives%

\begin{equation}
H = 2 {\bm v}^{T} \int_{0}^{\infty}e^{-2L_{k}t} dt~ L {\bm v}.
\end{equation}

\noindent According to the theory of spectral decomposition, $e^{-2L_{k}t} = \sum_{i=2}^{N} e^{-2\lambda_{i}} u_{i} u_{i}^{T}$, where $\lambda_{i}$ and
$u_{i}$ $(i = 2,3,...,N)$ are the eigenvalues of $L_{k}$ from the smallest to the largest (except for $\lambda_{1}= 0$) and $u_i$ their corresponding eigenvectors. Furthermore, as $L_{k}$ is positive semidefinite, all eigenvalues of $L_{k}$ are non-negative. Actually, except for $\lambda_{1}= 0$, the rest of eigenvalues are all positive so that the integral $\int_{0}^{\infty}e^{-2\lambda_{i}t}\mathrm{d}t$ is convergent. These conclusions combined lead to%

\begin{equation}
H = 2 {\bm v}^{T} \int_{0}^{\infty}\sum_{i=2}^{N} e^{-2\lambda_{i}t} u_{i}
u_{i}^{T} dt ~ L {\bm v}= 2 {\bm v}^{T} \sum_{i=2}^{N} \int_{0}^{\infty
}e^{-2\lambda_{i}t}dt ~ u_{i} u_{i}^{T}L {\bm v}={\bm v}^{T} \left(
\sum_{i=2}^{N} \frac{1}{\lambda_{i}}u_{i} u_{i}^{T}\right)  L {\bm v}.
\end{equation}

Similarly, the first derivative of $H$ is%

\begin{equation}
\frac{dH(k_{1})}{dk_{1}} = 2 {\bm v}^{T} \int_{0}^{\infty}\frac{d}{dk_{1}%
}(e^{-2[k_{1}(L_{1}-L_{2})+L_{2}]t}) dt~ L {\bm v} = -{\bm v}^{T}\left(
\sum_{i=2}^{N}\frac{1}{\lambda_{i}^{2}}u_{i}u_{i}^{T}\right)  (L_{1}%
-L_{2})L{\bm v},
\end{equation}

\noindent and the second derivative of $H$ is%

\begin{equation}
\frac{d^{2}H(k_{1})}{dk_{1}^{2}} = -4{\bm v}^{T} \int_{0}^{\infty}\frac
{d}{dk_{1}}(e^{-2L_{k}t})t dt~ (L_{1}-L_{2})L {\bm v} =2{\bm v}^{T}\left(
\sum_{i=2}^{N}\frac{1}{\lambda_{i}^{3}}u_{i}u_{i}^{T}\right)  (L_{1}%
-L_{2})^{2}L{\bm v}.
\end{equation}

\subsubsection*{S3. Further explanation of the choice of $\bm{v}$.}

The choice of $v$ has a direct effect on whether $H$ has the minimum value in (0,1). The key point lies in the sign of $\frac{dH(k_{1})}{dk_{1}}\bigg{|}_{k_{1}=\varepsilon}$.

Let $Q = \sum_{i=2}^{N}\frac{1}{\lambda_{i}^{2}}u_{i}u_{i}^{T}(L_{1}-L_{2})L
$. $Q$ is symmetric so that its eigenvectors compose an orthogonal basis.
Denote its eigenvalues and eigenvectors by $\{\mu_{i}\}_{i=1}^{N}$ and
$\{q_{i}\}_{i=1}^{N}$. $v$ can be rewritten as the linear combination of
$\{q_{i}\}_{i=1}^{N}$, i.e., ${\bm v}=\sum_{i} \beta_{i}q_{i}$. $\frac
{dH(k_{1})}{dk_{1}}$ can be expressed as:
\begin{equation}
\frac{dH(k_{1})}{dk_{1}} =-{\bm v}^{T}Q{\bm v} =-(\sum_{i} \beta_{i}q_{i}%
^{T})Q(\sum_{j} \beta_{j}q_{j})=-\sum_{i,j}\beta_{i}\beta_{j}q_{i}^{T}%
Qq_{j}=-\sum_{i,j}\mu_{j}\beta_{i}\beta_{j}q_{i}^{T}q_{j} =-\sum_{i}\mu
_{i}\beta_{i}^{2}. \label{ex}%
\end{equation}

\noindent That is, for the given $k_{1}=\varepsilon$, we can distribute
$\{\beta_{i}\}$ to ensure that Equation (\ref{ex}) is less than 0 and $H$ has
the minimum point in (0,1).

\newpage
\subsubsection*{S4. More curves corresponding to different mutualism networks}

\begin{figure}[h]
\centering
\includegraphics[scale=0.42]{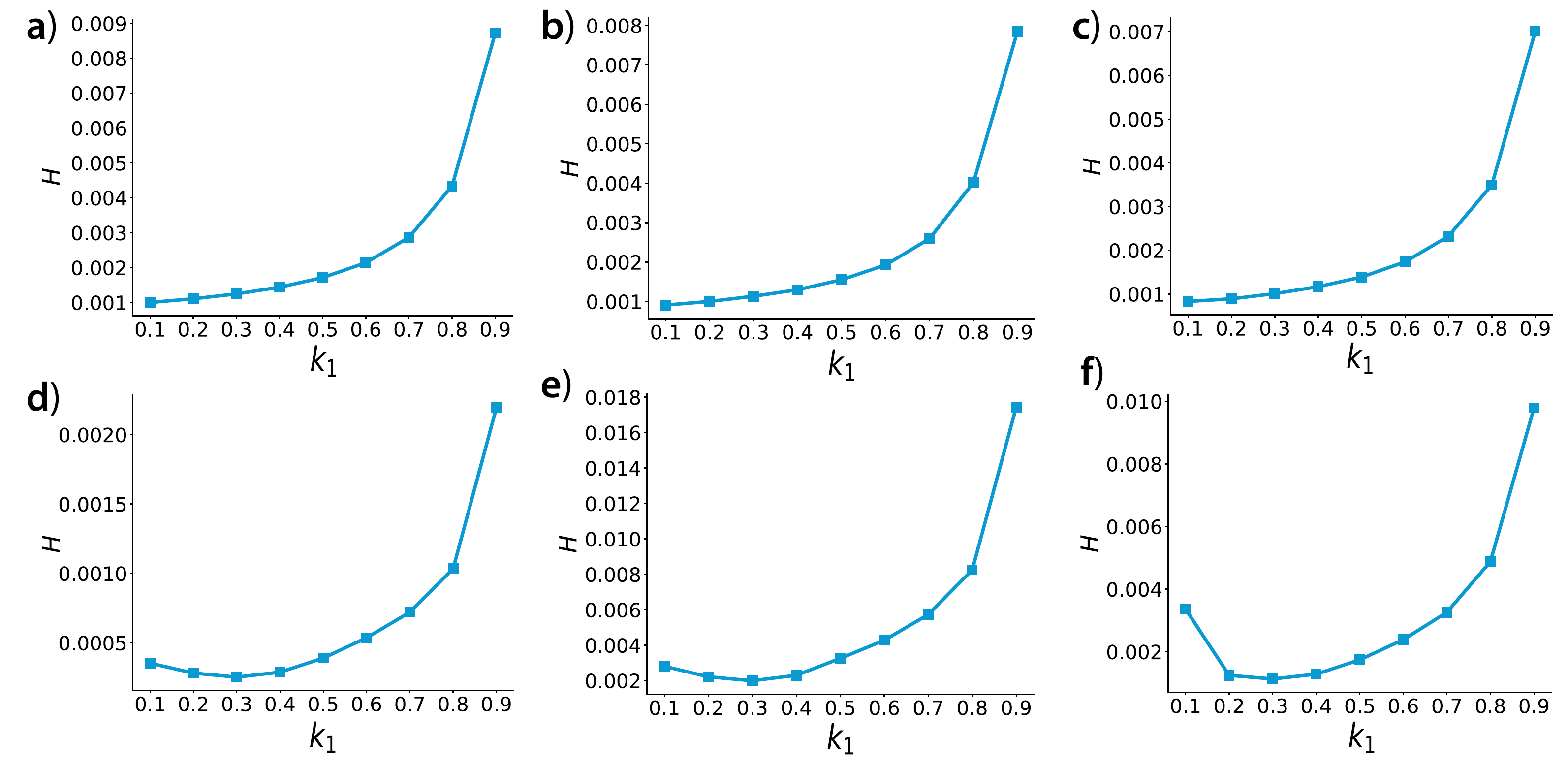}\caption{{\protect\footnotesize Coherency metric curve corresponding to six different mutualism networks. We set $B_{i}=B=0.01$, $r_{i}=r=0.01$, $C_{i}=C=5$, $G_{i}=G=1$, $D_{i}=D=5$, $E_{i}=E=0.9$ and $H_{i}=H=0.1$. ${\bm v}$ obeys a normal distribution $N(0,0.01)$. (a) 11 plants and 38 pollinators with 106 mutualistic interactions (network MPL08). (b) 14 plants and 13 pollinators with 52 mutualistic
interactions (network MPL11). (c) 7 plants and 33 pollinators with 65 mutualistic interactions (network MPL32). (d) 10 plants and 12 pollinators with 30 mutualistic interactions (network MPL36). (e) 16 plants and 44 pollinators with 278 mutualistic interactions (network MPL46). (f) 14 plants and 35 pollinators with 86 mutualistic interactions (network MPL50).}}%
\label{more_curves}%
\end{figure}

Fig.~\ref{more_curves} shows the coherency metric curve corresponding to six different mutualism networks. State change curves have similar patterns, which are omitted here. The result indicates that both patterns of coherency metric and state change are common in nature.

\bibliographystyle{unsrt}
\bibliography{manuscript}

\end{document}